\documentclass[reprint,twocolumn,longbibliography,superscriptaddress]{revtex4-2}
\usepackage{graphicx}
\usepackage{dcolumn}
\usepackage{bm}
\usepackage{inputenc}
\usepackage{epsfig}
\usepackage{float}
\usepackage{dcolumn}   
\usepackage{array}
\usepackage{amssymb}   
\usepackage{amsmath}
\usepackage{amstext}
\usepackage{mathtools}
\usepackage{threeparttable}
\usepackage{siunitx} 
\usepackage{lipsum}  
\usepackage{mathrsfs}
\usepackage{physics}
\usepackage{xcolor}
\usepackage[draft]{todonotes}
\usepackage{float}
\usepackage{lineno}

\usepackage [colorlinks=true,allcolors=blue]{hyperref}

\def\@element#1#2\@nil{%
  #1%
 \if\relax#2\relax\else\MakeLowercase{#2}\fi}
\makeatother
\begin{document}
\preprint{APS/123-QED}
\title{Unlocking Static Polarization and Strain Density Waves in Perovskites by Softening a Hidden Antiferrodistortive Tilt Gradient Mode}


\author{Yajun Zhang}
\affiliation{Key Laboratory of Mechanics on Disaster and Environment in Western China Attached to The Ministry of Education of China, Lanzhou University, Lanzhou 730000, Gansu, China}%
\affiliation{Department of Mechanics and Engineering Science, College of Civil Engineering and Mechanics, Lanzhou University, Lanzhou 730000, Gansu, China}%
\affiliation{School of Mechanical and Aerospace Engineering, Nanyang Technological University, Singapore 639798, Singapore}%

\author{Devesh R. Kripalani}
\affiliation{School of Mechanical and Aerospace Engineering, Nanyang Technological University, Singapore 639798, Singapore}%
\author{Xu He}
\affiliation{Theoretical Materials Physics, Q-MAT, Universit\'e de Li\`ege, B-4000 Li\`ege, Belgium}%

\author{Konstantin Shapovalov}
\affiliation{Theoretical Materials Physics, Q-MAT, Universit\'e de Li\`ege, B-4000 Li\`ege, Belgium}%

\author{Jiyuan Yang}
\affiliation{Department of Physics, School of Science, Westlake University, Hangzhou 310024, China}%
\author{Hongjian Zhao}
\affiliation{Key Laboratory of Material Simulation Methods and Software of Ministry of Education, College of Physics,
Jilin University, Changchun 130012, China}%

\author{Shi Liu}
\affiliation{Department of Physics, School of Science, Westlake University, Hangzhou 310024, China}%

\author{Huadong Yong}
\email{yonghd@lzu.edu.cn} 
\affiliation{Key Laboratory of Mechanics on Disaster and Environment in Western China Attached to The Ministry of Education of China, Lanzhou University, Lanzhou 730000, Gansu, China}%
\affiliation{Department of Mechanics and Engineering Science, College of Civil Engineering and Mechanics, Lanzhou University, Lanzhou 730000, Gansu, China}%

\author{Xingyi Zhang}
\email{zhangxingyi@lzu.edu.cn} 
\affiliation{Key Laboratory of Mechanics on Disaster and Environment in Western China Attached to The Ministry of Education of China, Lanzhou University, Lanzhou 730000, Gansu, China}%
\affiliation{Department of Mechanics and Engineering Science, College of Civil Engineering and Mechanics, Lanzhou University, Lanzhou 730000, Gansu, China}%

\author{Jie Wang }
\affiliation{Department of Engineering Mechanics, Zhejiang University, 38 Zheda Road, Hangzhou 310027, China}%

\author{Kun Zhou}
\email{kzhou@ntu.edu.sg} 
\affiliation{School of Mechanical and Aerospace Engineering, Nanyang Technological University, Singapore 639798, Singapore}%

\author{ Philippe Ghosez}
\affiliation{Theoretical Materials Physics, Q-MAT, Universit\'e de Li\`ege, B-4000 Li\`ege, Belgium}%

\begin{abstract}
Spin density waves (SDWs) represent a fundamental paradigm of spatially modulated order in condensed matter systems, yet their electrical and mechanical analogues--polarization and strain density waves (PDWs and StDWs)--have remained elusive as equilibrium phases. Here, we introduce a general, symmetry-driven strategy to unlock static PDWs and StDWs in perovskites SrTiO$_3$ and SrMnO$_3$. Using first-principles calculations, we uncover a previously overlooked soft antiferrodistortive tilt gradient mode at small-$q$ wavevector in the phonon dispersion of their presumed $Ima2$ ground state under moderate tensile strain. Group-theory analysis reveals that a hard polar-acoustic phonon, which intrinsically carries PDWs and StDWs, is improperly destabilized by a trilinear coupling with this modulated tilt mode and an inherently uniform tilt mode. This interaction drives a structural transition from the $Ima2$ phase to a novel lower-energy $Pmn2_1$ phase that hosts long-range-ordered PDWs and StDWs. Strikingly, the engineered StDWs in SrMnO$_3$ activate an electrically tunable SDW via the flexomagnetic effect. These discoveries fundamentally revise the strain-phase diagrams of prototypical perovskites and establish a unified phonon-engineering framework that links modulated phonon instabilities to targeted density-wave order, offering new pathways for designing advanced electromechanical and magnetoelectric functionalities.

\end{abstract}

\maketitle


\textit{Introduction--}Ferroic materials (ferroelectric (FE), ferromagnetic, ferroelastic) provide fertile ground for emergent quantum phenomena and advanced functionalities \cite{salje1990phase,devonshire1954theory,schmid1994multi,salje2012ferroelastic,martin2016thin,moya2014caloric,eerenstein2006multiferroic,fiebig2016evolution,junquera2023topological,schiaffino2017macroscopic}, owing to the strong coupling between lattice, charge, and spin degrees of freedom. Over the past decades, spatially inhomogeneous magnetization, polarization, and strain have emerged as central drivers of exotic behaviors, including unconventional superconductivity \cite{dai2012magnetism,hameed2022enhanced,enderlein2020superconductivity}, enhanced piezoelectric responses \cite{yao2025fluctuating,kutnjak2006giant,ahart2008origin,han2023tuning}, flexoelectricity \cite{zubko2013flexoelectric,lee2011giant,cross2006flexoelectric,zubko2007strain,biancoli2015breaking}, flexophotovoltaic effect \cite{yang2018flexo,shu2020photoflexoelectric}, and flexomagnetism \cite{makushko2022flexomagnetism,belyaev2020strain}. The creation and control of such inhomogeneous states has therefore become a key frontier in modern condensed matter physics.

A paradigmatic example of such inhomogeneous order is the spin density waves (SDWs) \cite{gruner1994dynamics}, characterized by periodic magnetization modulations that break symmetry, underpin exotic states like unconventional superconductivity \cite{dai2012magnetism}, and enable non-volatile memory \cite{chen2022resistive}. By contrast, realizing its electrical and mechanical counterparts, namely static polarization and strain density waves (PDWs and StDWs) remains an outstanding challenge in ferroic systems. Specifically, we define a StDW as a spontaneous, static modulation exhibiting long-range periodic strain field that parallels charge, spin, and polarization density waves. A major advance came recently with the observation of PDWs in terahertz-driven SrTiO$_3$ (STO) \cite{orenstein2025observation}, mediated by a transiently softened polar-acoustic phonon. The experiment confirmed long-standing speculation of fluctuating order in STO \cite{muller1991indication,rowley2014ferroelectric}. This non-equilibrium realization, however, raises a fundamental and broadly relevant question: can polar-acoustic phonons be intrinsically softened to stabilize static PDWs and StDWs in equilibrium ferroic materials?

Intriguingly, recent experiments on tensile-strained STO membranes have revealed anomalous quantum behavior near the paraelectric–FE phase boundary \cite{li2025classical}, strongly suggesting the presence of hidden emergent orders not captured by existing theories. Despite extensive theoretical and experimental investigations into the strain–phase diagrams of STO thin films \cite{he2005strain,warusawithana2009ferroelectric,zhang2024intrinsic,antons2005tunability,vasudevarao2006multiferroic,berger2011band,angsten2018epitaxial,pertsev2000phase,li2006phase,sheng2010phase,he2022structural}, direct evidence for softened polar-acoustic modes, the key microscopic precursors of PDWs and StDWs, has remained absent. With recent breakthroughs in atomic-precision strain engineering and freestanding ferroic membranes \cite{han2023tuning,li2025classical,han2024freestanding}, it is now timely to revisit tensile strain–phase diagrams of STO from an atomic-scale perspective to explore hidden lattice instabilities capable of stabilizing equilibrium modulated ferroic states.

In this Letter, we propose a generic phonon-mediated mechanism for engineering hidden polarization and strain density waves, enabled by a soft antiferrodistortive tilt instability, in stretched STO and SrMnO$_3$ (SMO) membranes. Our lattice-dynamics analysis reveals a hidden modulated longitudinal tilt instability in the presumed $Ima2$ ground state of tensile-strained STO and SMO. This modulated tilt mode trilinearly couples with the intrinsically uniform tilt mode and a hard polar-acoustic mode $\Sigma_2$ with modulated PDWs and StDWs, driving a structural transition to a hitherto unknown, lower-energy $Pmn2_1$ phase that exhibits spontaneous periodic PDWs and StDWs. Remarkably, in SMO, the engineered StDWs directly activate an emergent SDW with modulated magnetization through the flexomagnetic effect, allowing for electrical tuning of flexomagnetism. More broadly, we reveal that octahedral tilts act as a universal filter, suppressing higher-order harmonics and enforcing clean, single-$q$ modulations, in contrast to the multi-$q$ stripe-like $a_1/a_2$ \SI{90}{\degree} domain structures of PbTiO$_3$ (PTO). Our work revises the strain-phase diagram of paradigmatic perovskites STO and SMO and establishes a generic paradigm for unlocking ferroic and antiferrodistortive density waves in functional materials.

\textit{Cross-scale theory of spontaneous PDWs and StDWs.--}The spontaneous emergence of long-range periodic order in polarization and strain constitutes a fundamental challenge. Here, we present a phonon engineering strategy to bridge microscopic lattice dynamics with continuum field theory and unlock long-range polar and elastic modulations in tensile strained STO and SMO. Computational details are provided in the Supplemental Material. \cite {{note1}}\nocite {kresse1993ab,blochl1994projector,perdew1981self,perdew2008restoring,liechtenstein1995density,guo2018strain,grosso2021prediction,monkhorst1976special,togo2015first,stokes2006isodisplace,fu2003first,resta1994macroscopic}.

\begin{figure}
		\centering\includegraphics[width=\columnwidth]{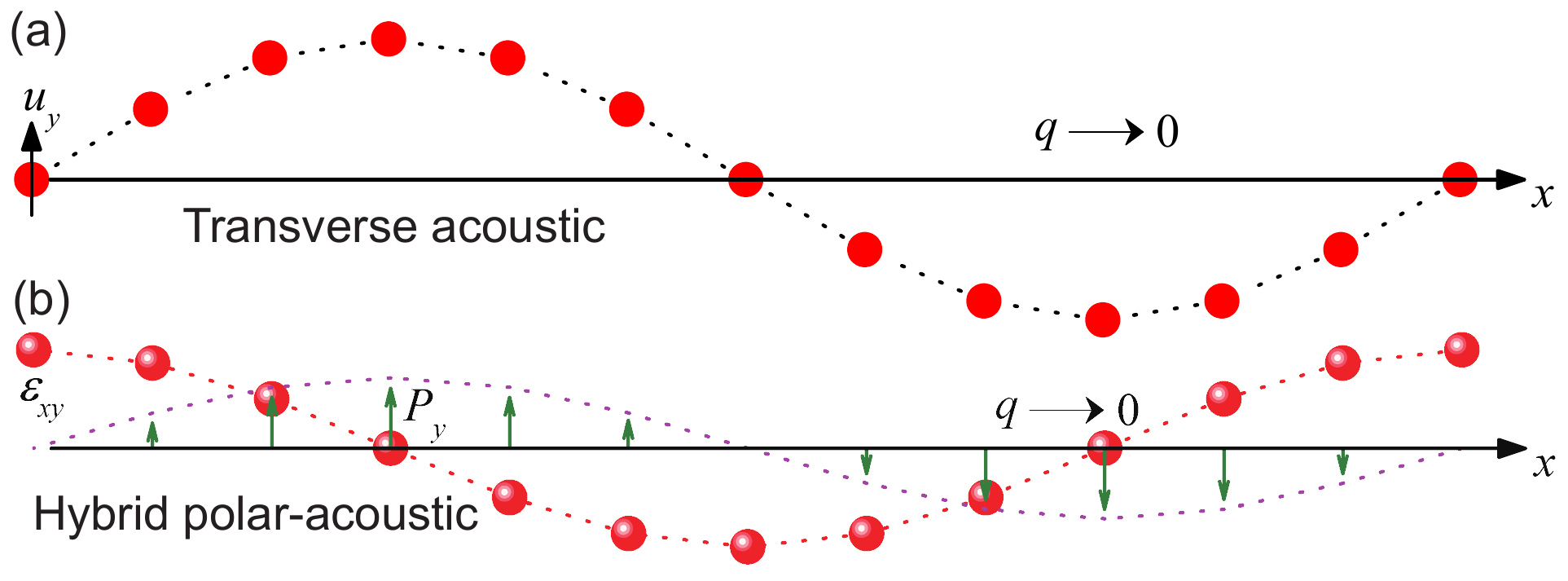}
    \caption{Schematic of (a) transverse acoustic phonon with modulated atomic displacements, and (b) hybrid polar-acoustic phonon coupling shear strain to polarization propagating along the $x$ axis. The red circles, red balls, and olive arrows denote the $y$-axis displacement, strain, and polarization, respectively.}
   \label{fig:Fig1}
\end{figure}

To establish a unified framework, we begin by considering the basic physics of acoustic phonons. A pure transverse acoustic (TA) phonon propagating along the $x$ axis [Fig. 1(a)] produces a sinusoidal displacement field $u_y(x) = A\sin(qx)$ \cite{cowley1976acoustic}, which directly induces a periodic shear strain field $\varepsilon_{xy}(x) = \frac{1}{2} \frac{\partial u_y}{\partial x} = \frac{A q}{2} \cos(q x)$.



In such a TA mode, polarization emerges indirectly as a secondary effect via symmetry breaking or flexoelectricity \cite{guzman2023lamellar}. A hybrid polar-acoustic phonon at a small wavevector $q$ is fundamentally different. As sketched in Fig. 1(b), such a mode directly couples shear strain to dipole formation, generating both modulated polarization $P_y(x) = P_0 \sin(qx + \phi)$ and strain fields from a single soft mode. 

This theoretical framework establishes a cross-scale foundation for spontaneous polar and elastic modulations. While such hybrid mode remains stable in bulk perovskites, we discover a previously overlooked unstable antiferrodistortive tilt gradient mode in tensile-strained perovskite membranes such as STO and SMO, which, through symmetry-allowed coupling, provides a generic mechanism to activate otherwise stable polar-acoustic modes, driving the formation of long-range ordered density waves (Fig. 2).

\begin{figure*}[t]
	\centering\includegraphics[width=0.95\textwidth]{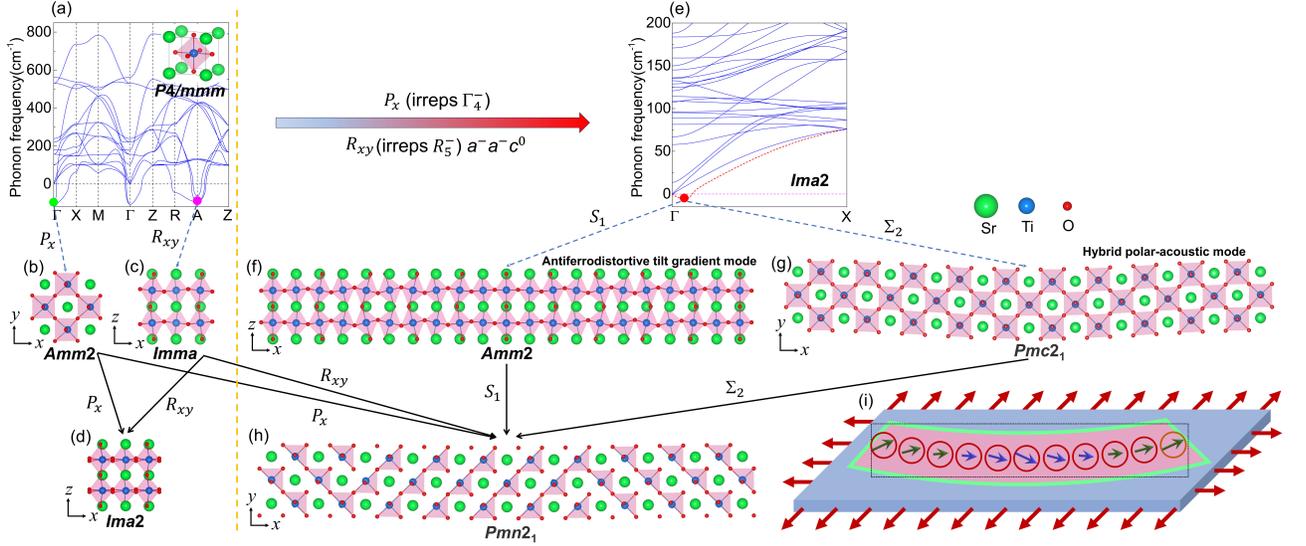}
    \caption{Atomic theory of spontaneous PDW and StDW. The $x$, $y$, $z$ axes correspond to $[110]$, $[1\bar{1}0]$, and $[001]$ of the pseudocubic phase. (a) Phonon dispersion of $P4/mmm$ STO under 1.5\% tensile strain; inset shows the corresponding atomic structure. Schematic representations of (b) in-plane polarization $P_x$, (c) out-of-phase tilt mode $R_{xy}$ (projected along $[1\bar{1}0]$, with rotation around $[110]$), and (d) $Ima2$ phase with $P_x$ and $R_{xy}$ distortions. (e) Phonon dispersion ($\Gamma$–X) of $Ima2$ STO under the same strain, revealing a hidden instability. (f) Modulated tilt mode $S_1$ and (g) hybrid polar-acoustic mode $\Sigma_2$ carrying PDW and StDW. (h) The $Pmn2_1$ ground state incorporating all four distortions ($P_x$, $R_{xy}$, $S_1$, $\Sigma_2$). (i) Schematic of biaxial tensile strain (brown arrows) inducing PDW and StDW in a freestanding membrane; the arrows in the circles represent the modulated total polarization vectors. The substrate-like layer is a conceptual reference to experimental setups using sacrificial layers, actual DFT calculations use periodic supercells without substrates. The green dashed line separates the conventional understanding (left) from our new findings (right) on the lowest-energy state of STO under moderate tensile strain.}
   \label{fig:Fig1}
\end{figure*}


\begin{figure}
		\centering\includegraphics[width=0.9\columnwidth]{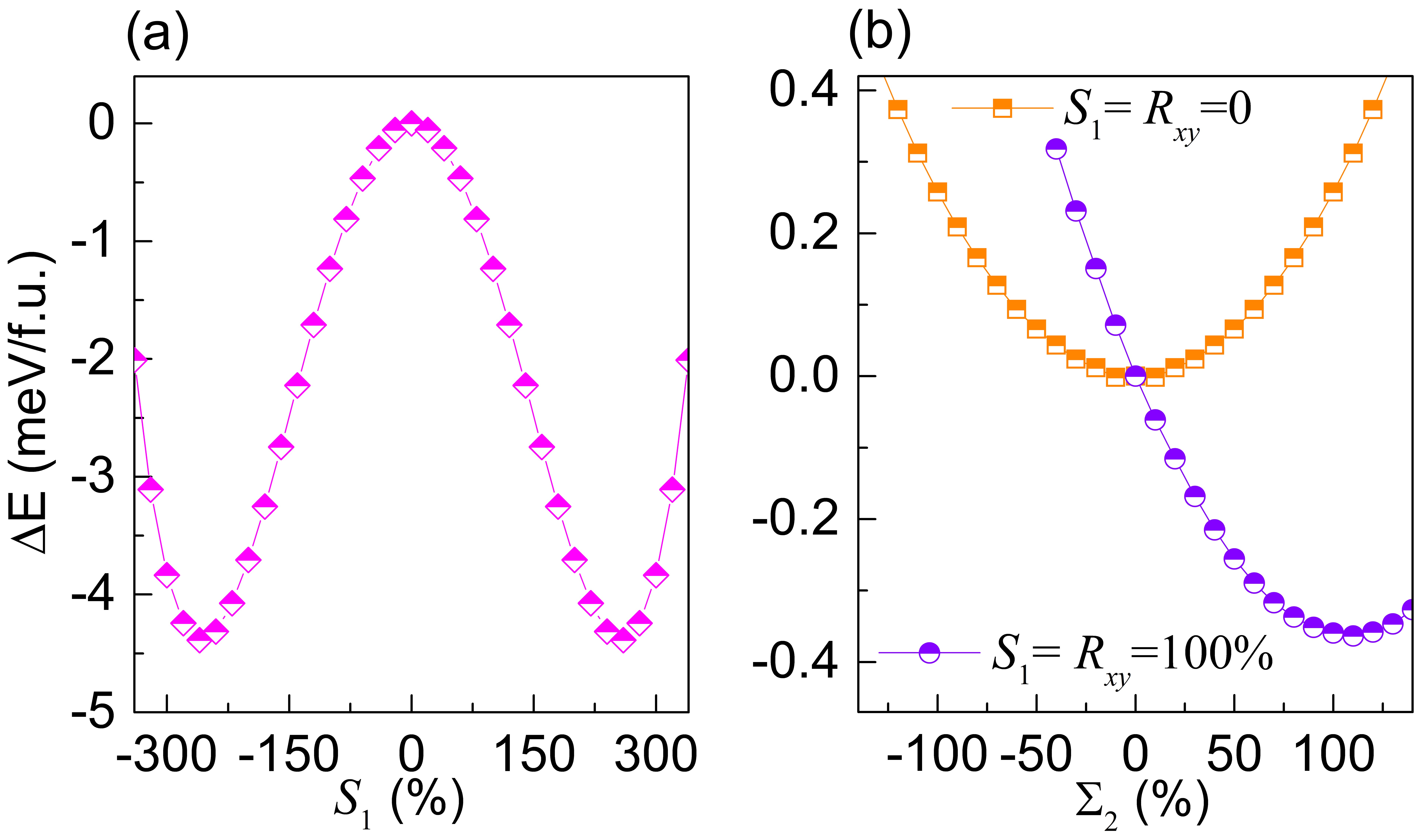}
    \caption{PESs of the (a) $S_{1}$ mode and the (b) $\Sigma_2$ mode without (orange curve) and with (blue curve) the fixing of $R_{xy}$ and $S_{1}$ modes. Here, 100\% denotes the amplitude of distortions in STO under 1.5\% tensile strain.}
   \label{fig:Fig1}
\end{figure}

\textit{Improper PDWs and StDWs in STO membranes.--}STO, a prototypical incipient ferroelectric known for its pronounced strain–phonon coupling, serves as a model system for exploring strain-induced ferroelectricity \cite{he2005strain,warusawithana2009ferroelectric,zhang2024intrinsic,antons2005tunability,vasudevarao2006multiferroic,berger2011band,angsten2018epitaxial,pertsev2000phase,li2006phase,sheng2010phase,he2022structural}. Extensive theoretical studies including Ginzburg–Landau–Devonshire phenomenology \cite{pertsev2000phase}, phase-field simulations \cite{li2006phase,sheng2010phase}, deep potential molecular dynamics \cite{he2022structural}, and first-principles calculations \cite{vasudevarao2006multiferroic,berger2011band,angsten2018epitaxial} have established a consistent phase transition for STO under tensile strain. Taking $1.5\%$ biaxial tensile strain as an example, the high-temperature reference $P4/mmm$ phase (the cubic structure under in-plane biaxial strain) exhibits two main lattice instabilities in its phonon spectrum shown in Fig. 2(a): unstable polar modes at the $\Gamma$ point corresponding to in-plane FE distortions $P_x$ [Fig. 2(b)], and unstable antiferrodistortive modes at the $A$ point $(1/2,1/2,1/2)$ corresponding to oxygen octahedral tilts. Under tensile strain, the $a^-a^-c^0$ tilt $R_{xy}$ [Fig. 2(c)] becomes significantly softer than the $a^0a^0c^-$ pattern, making it the dominant tilt instability. Condensing the $P_x$ and the octahedral tilt $R_{xy}$ leads to the widely accepted FE $Ima2$ ground state [Fig. 2(d)] \cite{vasudevarao2006multiferroic,berger2011band,angsten2018epitaxial}.

Contrary to this consensus, we discover a soft phonon branch near the $\Gamma$ point in the established $Ima2$ ground state at $1.5\%$ tensile strain [Fig. 2(e)], indicating a hitherto overlooked instability toward a distinct low-energy modulated phase. Symmetry analysis identifies the unstable eigenmode as mixed modes combining a modulated longitudinal tilt mode $S_1$ (Irrep $S_1$) and a hybrid polar-acoustic mode $\Sigma_2$ (Irrep $\Sigma_2$). As schematically depicted in Fig. 2(f), the $S_{1}$ component, with $Amm2$ symmetry, generates a longitudinal tilt modulated at wavevector $q$ = ($n-1/2n$, $n-1/2n$, 0) within a $\sqrt{2}n \times \sqrt{2} \times 2$ supercell. The $\Sigma_2$ mode ($Pmc2_1$ symmetry), as shown in Fig. 2(g), induces modulated displacement and polarization fields along the $x$ axis at $q$ = (1/2$n$, 1/2$n$, 0). The condensation of these modes and subsequent full structural relaxation within a $\sqrt{2}n \times \sqrt{2} \times 2$ supercell under $1.5\%$ tensile strain consistently yields a new $Pmn2_1$ phase [Fig. 2(h)], structurally distinct and energetically favored (Fig. S1 \cite{note1}) over the presumed $Ima2$ ground state.


To resolve the stabilization mechanism of this $Pmn2_1$ phase, we employ group theory symmetry analysis to untangle the symmetry relationships among key distortions ($P_x$, $R_{xy}$, $S_1$, and $\Sigma_2$). The potential energy surfaces (PESs) as a function of the normalized mode amplitudes [Fig. 3 and Fig. S2 \cite{note1}] show that the $P_x$, $R_{xy}$, and $S_1$ modes exhibit double-well potentials, confirming their roles as primary order parameters, whereas the $\Sigma_2$ mode displays a single-well potential [Fig. 3(b)], indicating its activation $via$ mode coupling. Remarkably, the $\Sigma_2$ mode only stabilizes at finite amplitude when both $R_{xy}$ and $S_1$ modes are simultaneously condensed [Fig. 3(b) and Figs. S3–S4 \cite{note1}]. The linear energy dependence near zero amplitude [Fig. 3(b) and Fig. S5 \cite{note1}] for these three modes provides direct symmetry-consistent evidence of a trilinear coupling term:

\begin{equation}
\begin{split}
\mathcal F = R_{xy} S_1 \Sigma_2.
\end{split}
\end{equation}

The resulting $Pmn2_1$ phase exhibits pronounced long-range spatial modulations. The octahedral tilt angle displays a cosinusoidal profile [Fig. 4(a)], manifesting as long-range tilt density waves (TDWs). Concurrently, atomic displacements along the $x$ axis follow a sinusoidal wave profile [Fig. 4(b)], generating a cosinusoidal shear strain wave [Fig. 4(c)] that constitutes a spontaneous StDW. Polarization mapping further reveals a modulated $P_y$ with identical periodicity [Fig. 4(d)]. The modulated $P_y$ and uniform $P_x$ form a coherent PDW, as schematically illustrated in Fig. 2(i). These findings fundamentally revise the tensile strain-phase diagram of STO, establish the $Pmn2_1$ phase as a versatile platform for hosting trilinear coupling--induced PDW and StDW, and provide a plausible mechanism for anomalous quantum behavior in strained membranes \cite{li2025classical}.

\begin{figure}
		\centering\includegraphics[width=\columnwidth]{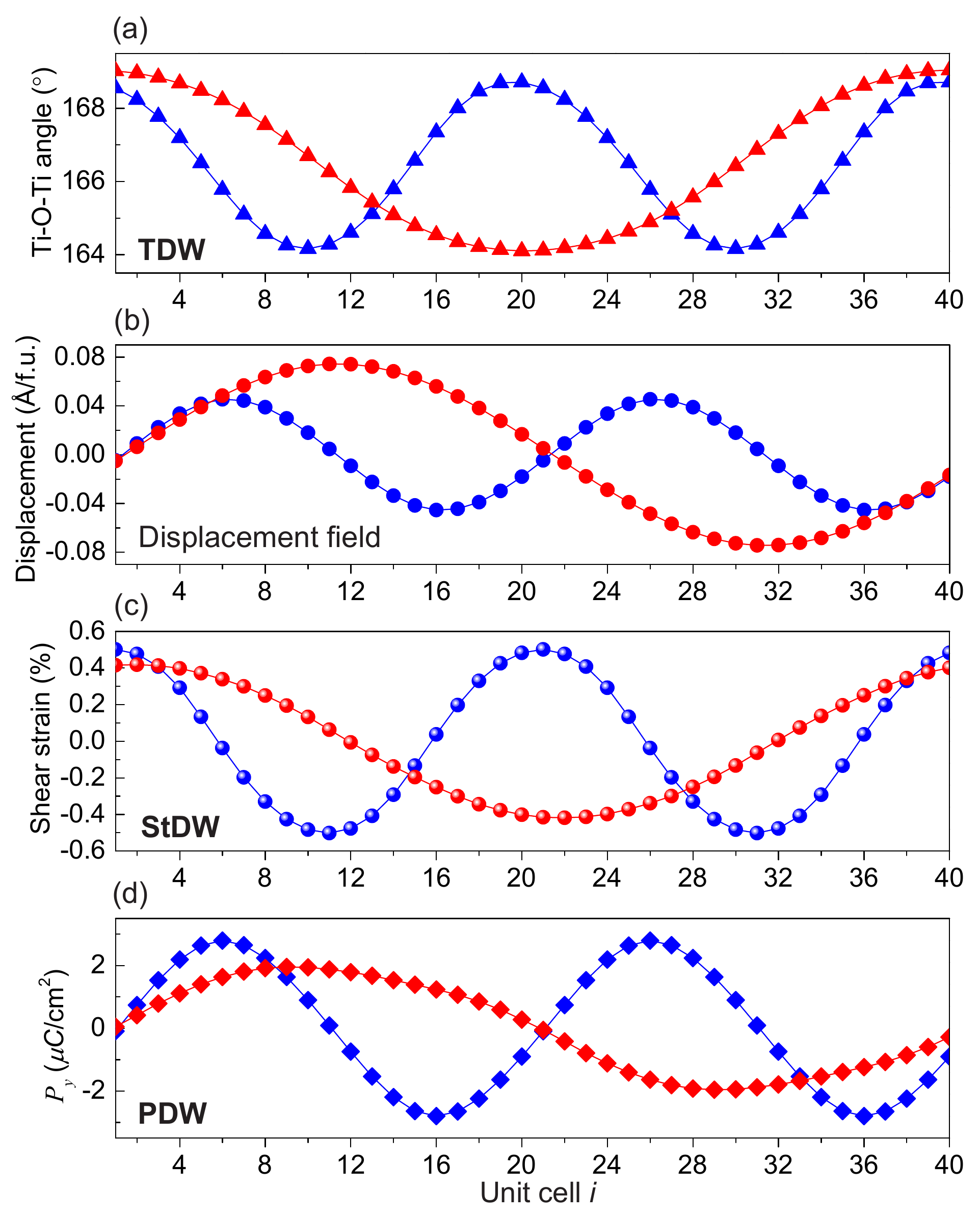}
    \caption{Improper PDW and StDW in tensile-strained STO. The (a) out-of-plane Ti-O-Ti bond angle, (b) displacement field, (c) strain field, and (d) $P_y$ along the $x$ axis. These fields are extracted from the $10\sqrt{2} \times \sqrt{2} \times 2$ (blue curves) supercell and $20\sqrt{2} \times \sqrt{2} \times 2$ (red curves) supercell of $Pmn2_1$ STO under 1.5\% tensile strain.}
   \label{fig:Fig1}
\end{figure}

\textit{Electrically switchable SDW in SMO.--}Flexo-type coupling effects have recently emerged as a critical research frontier, owing to their great potential for electromechanical and magnetomechanical technologies \cite{zubko2013flexoelectric,cross2006flexoelectric,lee2011giant,makushko2022flexomagnetism,belyaev2020strain}. Although pioneering studies demonstrated flexomagnetic effects \cite{makushko2022flexomagnetism,belyaev2020strain}, their implementation might be hindered by the difficulty of generating deterministic, nanoscale strain gradients via conventional bending methods.

Against this backdrop, we overcome this limitation by leveraging spontaneous StDW in perovskite oxides. While prior first-principles studies consistently identified a tensile strain stabilized $Ima2$ ground state in SMO \cite{lee2010epitaxial,cazorla2023giant,guo2018strain}, we discover that moderate strain stabilizes a modulated $Pmn2_1$ phase with coexisting PDW and StDW [Fig. 5(a) and S6 \cite{note1}], mirroring the behavior in STO. Crucially, this spontaneously generated StDW serves as an ideal built-in platform for producing nanoscale strain gradients, enabling efficient generation of SDW via the flexomagnetic effect.

Within a 3\% tensile strained SMO thin film modeled in an $8\sqrt{2} \times \sqrt{2} \times 2$ supercell, we demonstrate StDW-mediated SDW formation. Figure 5(b) illustrates the sinusoidal strain modulation and cosinusoidal polarization $P_y$ distribution, analogous to that observed in STO. These modulations directly induce periodic SDW [Fig. 5(c)] with magnetization modulations along the $x$ axis, the hallmark of the flexomagnetic effect. The SDW period is exactly half that of the strain wave, arising from equivalent magnetization suppression under both positive and negative shear strains (Fig. S7 \cite{note1}).

The coexistence of electrically tunable in-plane polarization $P_x$ and SDW enables direct electric-field control of magnetization. As quantified in Fig. 5(c), the magnetization difference depends on the amplitude and direction of in-plane electric field along the $x$ axis: a field antiparallel to $P_x$ enhances the SDW amplitude, whereas a parallel field suppresses it. Beyond a critical field strength, the modulated magnetization is fully quenched [Fig. 5(c)], demonstrating nonvolatile switching off of the StDW-mediated SDW. We thus establish amplitude modulation and non-volatile off switching of StDW-mediated SDW and flexomagnetism via electric fields alone, which offers a direct pathway toward ultra-low-power, voltage-controlled spintronic logic and memory devices.

\textit{PDW/StDW state $vs$ \SI{90}{\degree} domain.--}The intriguing PDWs and StDWs in modulated long-period STO and SMO stand in stark contrast to those in classical ferroelectrics such as PTO, which under tensile strain form a long-period $a_1/a_2$ $90^\circ$ domain structure characterized by flattened domains and sharp boundaries~\cite{li2017thickness,wang2020converse}.

\begin{figure} [t]
	\centering\includegraphics[width=\columnwidth]{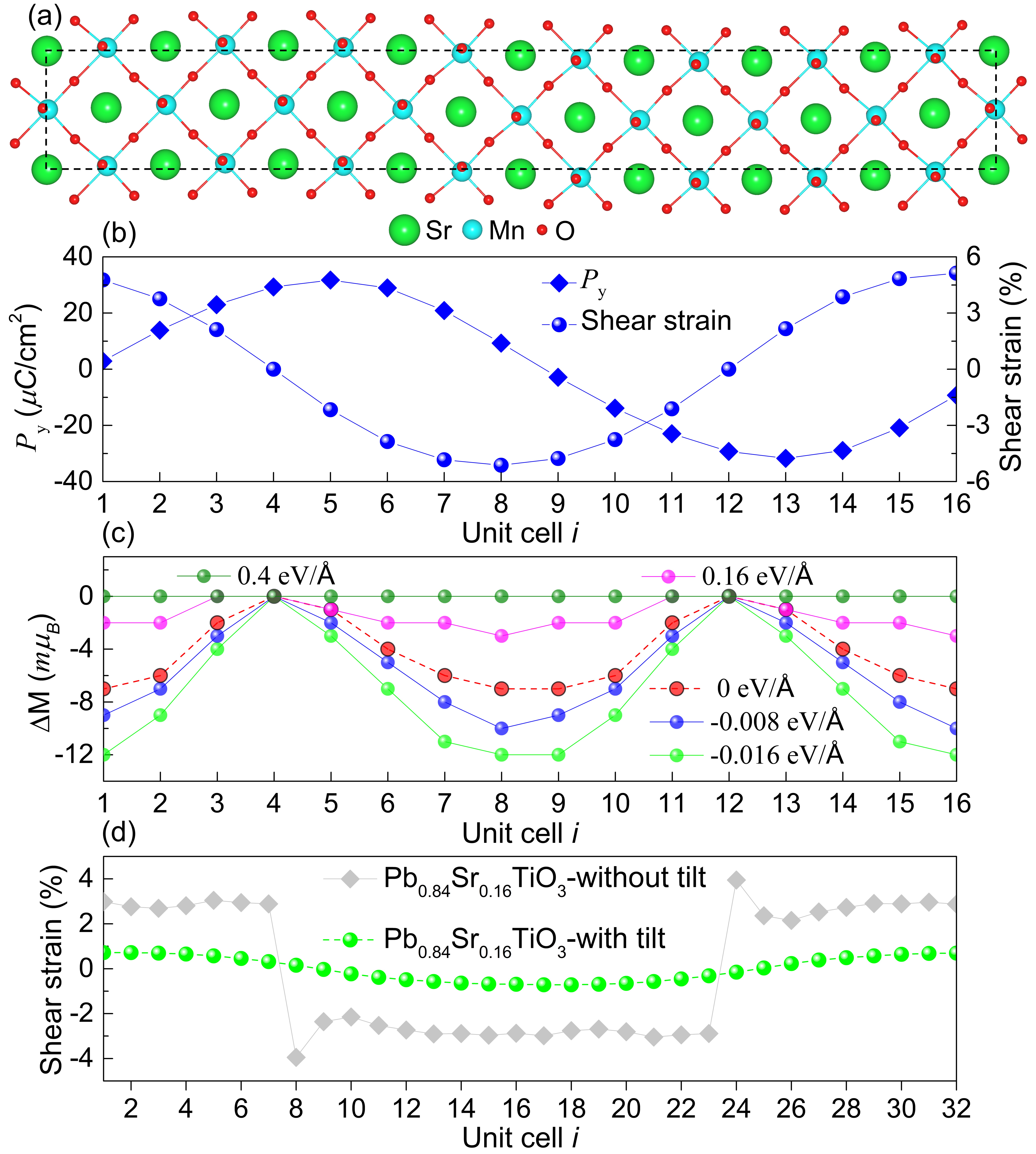}
    \caption{Electrical and chemical control of density waves. (a) Atomic structure of the low-energy $Pmn2_1$ phase in $8\sqrt{2} \times \sqrt{2} \times 2$ SMO supercell under 3\% tensile strain. (b) Corresponding global strain field and distribution of $P_y$ along the $x$ axis. (c) The spatial distribution of magnetization deviation along the $x$ axis under different in-plane electric field conditions. (d) Comparison of modulated strain field for Pb$_{0.84}$Sr$_{0.16}$TiO$_3$ grown on the DyScO$_3$ substrate without and with the involving of tilt distortions, highlighting the critical role of tilt distortions in suppressing higher-order harmonics and stabilizing single-$q$ dominant StDW.}
    \label{fig:Fig4:JT_coupling}
\end{figure}

\begin{table*} [t]
\centering\renewcommand\arraystretch{1.3}
\caption{Lattice distortions in the modulated FE phase of tensile-strained STO (1.5\%), SMO (3\%), and BFO (4.5\%) membranes, relative to the pseudocubic phase. All values are dimensionless, representing normalized mode amplitudes with respect to the pseudocubic structure. We also list the distortions of PTO and Sr-doped PTO films grown on DyScO$_3$ substrate.}
\label{tab:Tab2_Fitting_Paras}
\setlength{\tabcolsep}{0.7 mm}{
\begin{tabular}{ccccccccc}  \toprule
Compound &  Space group & $P_{x}$ & $R_{xy}$ & $\Sigma_2$, $_{q = 1/2n, 1/2n, 0}$ &$\Sigma_2$, $_{q = 3/2n, 3/2n, 0}$ & $\Sigma_2$, $_{q = 5/2n, 5/2n, 0}$ & $\Sigma_2$, $_{q = 7/2n, 7/2n, 0}$  \\	 
	 \hline 
STO & $Pmn2_1$ & 0.122 & 0.278 & 0.117 &  0.001  & 0.003 & 0.001   \\
SMO & $Pmn2_1$ & 0.143 & 0.205 & 0.809  & 0.010 & 0.008 &  0.001 \\
BFO  & $Pc$ & 1.235 & 0.633 & 0.211 &  0.002  &0.001 & 0.001   \\
PTO & $Pmc2_1$ & 0.501 & 0 & 1.362 &  0.192  & 0.069 &  0.036   \\  
Pb$_{0.84}$Sr$_{0.16}$TiO$_3$ & $Pmn2_1$ & 0.580 & 0.21 & 0.175 &  0.005  & 0 &  0   \\
 \hline   

 \hline  
\end{tabular}}
\end{table*}

To reconcile these distinct behaviors and establish a unified microscopic description of polarization and strain fields in tensile-strained ferroelectrics, we performed a systematic analysis of the $\Sigma_2$ mode harmonics across these systems. As summarized in Table 1, multiple $\Sigma_2$ modes exist relative to the cubic parent structure. In a $\sqrt{2}n \times \sqrt{2} \times 2$ supercell, each $\Sigma_2$ mode corresponds to a specific wavevector $q$ = ($m$/2$n$, $m$/2$n$, 0), where $m$ is an odd integer smaller than $n$. The mode with $q$ = (1/2$n$, 1/2$n$, 0) possesses the largest amplitude. The key distinction lies in the strength of the higher-order harmonic modes: STO and SMO exhibit negligible higher-order harmonics, preserving a single-$q$ modulation profile. A similar trend is observed in BFO, where higher-order harmonics are negligible (Table 1), leading to well-ordered, single-$q$ dominant PDWs and StDWs along $x$ axis (Fig. S8 \cite{note1}). In contrast, PTO exhibits strong higher-order harmonics, and the condensation of multiple $\Sigma_2$ modes gives rise to the anharmonic, stripe-like waveforms shown in Fig. S9 \cite{note1}.

Under tensile strain, a fundamental microscopic distinction lies in the presence or absence of octahedral tilts, which are inherent in STO, SMO, and BFO but absent in PTO. To clarify the effect of octahedral tilts, we investigate the strain field in Sr-doped PTO grown on a DyScO$_3$ substrate. As illustrated in Fig. S10 \cite{note1}, Sr doping systematically introduces octahedral tilts. If these tilts are neglected in the modeling, multi-$q$ condensation persists, leading to a stripe-like strain and displacement profile within the $a_1/a_2$ domain structure [Fig. 5(d)]. However, when the Sr-induced tilts are explicitly included, the higher-order harmonics are strongly suppressed (Table 1), driving the system from a multi-$q$ regime toward a nearly single-$q$ dominant state [Fig. 5(d)]. Our findings thus establish octahedral engineering as a powerful strategy for tuning density wave states and other functional properties in perovskite oxides $via$ selective control of anharmonic mode coupling.

To conclude, the pioneering work of Orenstein et al. provided the compelling observation of PDWs in STO under ultrafast terahertz excitation, revealing the intrinsically dynamical, nonequilibrium character mediated by flexoelectric coupling. Inspired by this landmark discovery, our work establishes a novel phonon engineering design principle that bridges atomic-scale lattice dynamics with equilibrium polarization and strain modulations in stretched perovskite membranes. We identify a previously overlooked soft antiferrodistortive tilt-gradient mode in tensile-strained STO and SMO. Through a symmetry-allowed improper trilinear coupling with a uniform tilt distortion and a hard polar-acoustic phonon, this hidden instability drives a transition from the presumed $Ima2$ ground state to a lower-energy, long-period $Pmn2_1$ phase hosting static PDW and StDW. In SMO, the resulting StDW generates intrinsic nanoscale strain gradients that create a flexomagnetic landscape, enabling direct electric-field control of SDWs and magnetic textures without external mechanical deformation. More broadly, we show that octahedral tilts suppress higher-order polar-acoustic harmonics, favoring single-$q$ dominant modulations, unlike the multi-$q$ domain patterns of classical ferroelectrics such as PTO. This behavior is further confirmed in BFO and Sr-doped PTO, demonstrating the generality of this mechanism across tilted perovskites. Consequently, these results establish a new paradigm for density-wave engineering. Instead of transient, field-driven excitations governed primarily by flexoelectric coupling, the modulated states identified here correspond to stable thermodynamic ground states arising from symmetry-guided trilinear phonon interactions and accessible through static tensile strain. While this manuscript was under review, an independent experimental study \cite{wang2026hidden} reported the observation of a hidden polar modulated phase in uniaxially tensile-strained STO, providing experimental confirmation of the strain-driven polar–acoustic instability mechanism identified in this work for both biaxially and uniaxially (see Fig. S11 \cite{note1}) tensile-strained STO. By demonstrating that these long-period modulations can serve as functional parent phases for electrical control of derivative orders such as SDW, our work opens a pathway toward static, electrically controllable multifunctional density-wave phases in complex oxides.

Acknowledgments--Y.Z. acknowledges the financial support received from the National Natural Science Foundation of China (Grant 12472150 and 12432007). Y.Z. also acknowledges the Center for Computational Science and Engineering of Lanzhou University for the computational support provided.
\bibliography{NNO-ref}

@article{he2005strain,
  title={Strain phase diagram and domain orientation in \rm {SrTiO$_3$} thin films},
  author={He, Feizhou and Wells, BO and Shapiro, SM},
  journal={Phys. Rev. Lett.},
  volume={94},
  number={17},
  pages={176101},
  year={2005},
  publisher={APS}
}

@article{hameed2022enhanced,
  title={Enhanced superconductivity and ferroelectric quantum criticality in plastically deformed strontium titanate},
  author={Hameed, Sajna and Pelc, Damjan and Anderson, Zachary W and Klein, Avraham and Spieker, RJ and Yue, L and Das, Bhaskar and Ramberger, Justin and Lukas, Marin and Liu, Yaohua and others},
  journal={Nat. Mater.},
  volume={21},
  number={1},
  pages={54--61},
  year={2022},
  publisher={Nature Publishing Group UK London}
}

@article{wang2026hidden,
  title={Hidden polar phase in the quantum paraelectric SrTiO3},
  author={Wang, Huaiyu Hugo and Flores, Ernesto and Stanton, Jade and Orenstein, Gal and Miedaner, Peter R and Foglia, Laura and Martinez, Maya and Reis, David A and Mankowsky, Roman and Sander, Mathias and others},
  journal={arXiv preprint arXiv:2603.12239},
  year={2026}
}

@article{ahart2008origin,
  title={Origin of morphotropic phase boundaries in ferroelectrics},
  author={Ahart, Muhtar and Somayazulu, Maddury and Cohen, Ronald E and Ganesh, P and Dera, Przemyslaw and Mao, Ho-kwang and Hemley, Russell J and Ren, Yang and Liermann, Peter and Wu, Zhigang},
  journal={Nature},
  volume={451},
  number={7178},
  pages={545--548},
  year={2008},
  publisher={Nature Publishing Group UK London}
}

@article{zubko2013flexoelectric,
  title={Flexoelectric effect in solids},
  author={Zubko, Pavlo and Catalan, Gustau and Tagantsev, Alexander K},
  journal={Annu. Rev. Mater. Res.},
  volume={43},
  number={1},
  pages={387--421},
  year={2013},
  publisher={Annual Reviews}
}

@article{guzman2023lamellar,
  title={Lamellar fluctuations melt ferroelectricity},
  author={Guzm{\'a}n-Verri, GG and Liang, CH and Littlewood, PB},
  journal={Phys. Rev. Lett.},
  volume={131},
  number={4},
  pages={046801},
  year={2023},
  publisher={APS}
}

@article{monkhorst1976special,
  title={Special points for Brillouin-zone integrations},
  author={Monkhorst, Hendrik J and Pack, James D},
  journal={Phys. Rev. B},
  volume={13},
  number={12},
  pages={5188},
  year={1976},
  publisher={APS}
}

@article{perdew2008restoring,
  title={Restoring the density-gradient expansion for exchange in solids and surfaces},
  author={Perdew, John P and Ruzsinszky, Adrienn and Csonka, G{\'a}bor I and Vydrov, Oleg A and Scuseria, Gustavo E and Constantin, Lucian A and Zhou, Xiaolan and Burke, Kieron},
  journal={Phys. Rev. Lett.},
  volume={100},
  number={13},
  pages={136406},
  year={2008},
  publisher={APS}
}

@article{fu2003first,
  title={First-Principles Determination of Electromechanical Responses of Solids under Finite Electric Fields},
  author={Fu, Huaxiang and Bellaiche, L},
  journal={Phys. Rev. Lett.},
  volume={91},
  number={5},
  pages={057601},
  year={2003},
  publisher={APS}
}

@article{resta1994macroscopic,
  title={Macroscopic polarization in crystalline dielectrics: the geometric phase approach},
  author={Resta, Raffaele},
  journal={Rev. Mod. Phys.},
  volume={66},
  number={3},
  pages={899},
  year={1994},
  publisher={APS}
}

@article{antons2005tunability,
  title={Tunability of the dielectric response of epitaxially strained \rm {SrTiO$_3$} from first principles},
  author={Antons, Armin and Neaton, JB and Rabe, Karin M and Vanderbilt, David},
  journal={Phys. Rev. B},
  volume={71},
  number={2},
  pages={024102},
  year={2005},
  publisher={APS}
}

@article{cross2006flexoelectric,
  title={Flexoelectric effects: Charge separation in insulating solids subjected to elastic strain gradients},
  author={Cross, L Eric},
  journal={J. Mater. Sci.},
  volume={41},
  pages={53--63},
  year={2006},
  publisher={Springer}
}

@article{lee2011giant,
  title={Giant flexoelectric effect in ferroelectric epitaxial thin films},
  author={Lee, Daesu and Yoon, A and Jang, SY and Yoon, J-G and Chung, J-S and Kim, M and Scott, JF and Noh, TW},
  journal={Phys. Rev. Lett.},
  volume={107},
  number={5},
  pages={057602},
  year={2011},
  publisher={APS}
}

@article{angsten2018epitaxial,
  title={Epitaxial phase diagrams of \rm {SrTiO$_3$}, \rm {CaTiO$_3$}, and \rm {SrHfO$_3$}: Computational investigation including the role of antiferrodistortive and A-site displacement modes},
  author={Angsten, Thomas and Asta, Mark},
  journal={Phys. Rev. B},
  volume={97},
  number={13},
  pages={134103},
  year={2018},
  publisher={APS}
}

@article{kutnjak2006giant,
  title={The giant electromechanical response in ferroelectric relaxors as a critical phenomenon},
  author={Kutnjak, Zdravko and Petzelt, Jan and Blinc, Robert},
  journal={Nature},
  volume={441},
  number={7096},
  pages={956--959},
  year={2006},
  publisher={Nature Publishing Group UK London}
}

@article{han2023tuning,
  title={Tuning piezoelectricity via thermal annealing at a freestanding ferroelectric membrane},
  author={Han, Lu and Yang, Xinrui and Lun, Yingzhuo and Guan, Yue and Huang, Futao and Wang, Shuhao and Yang, Jiangfeng and Gu, Chenyi and Gu, Zheng-Bin and Liu, Lisha and others},
  journal={Nano Lett.},
  volume={23},
  number={7},
  pages={2808--2815},
  year={2023},
  publisher={ACS Publications}
}

@article{pertsev2000phase,
  title={Phase transitions and strain-induced ferroelectricity in \rm {SrTiO$_3$} epitaxial thin films},
  author={Pertsev, NA and Tagantsev, AK and Setter, N},
  journal={Phys. Rev. B},
  volume={61},
  number={2},
  pages={R825},
  year={2000},
  publisher={APS}
}

@article{he2022structural,
  title={Structural phase transitions in \rm {SrTiO$_3$} from deep potential molecular dynamics},
  author={He, Ri and Wu, Hongyu and Zhang, Linfeng and Wang, Xiaoxu and Fu, Fangjia and Liu, Shi and Zhong, Zhicheng},
  journal={Phys. Rev. B},
  volume={105},
  number={6},
  pages={064104},
  year={2022},
  publisher={APS}
}

@article{sheng2010phase,
  title={Phase transitions and domain stabilities in biaxially strained (001) \rm {SrTiO$_3$} epitaxial thin films},
  author={Sheng, Guang and Li, YL and Zhang, JX and Choudhury, S and Jia, QX and Gopalan, Venkatraman and Schlom, Darrell G and Liu, ZK and Chen, LQ},
  journal={J. Appl. Phys.},
  volume={108},
  number={8},
  year={2010},
  publisher={AIP Publishing}
}

@article{warusawithana2009ferroelectric,
  title={A ferroelectric oxide made directly on silicon},
  author={Warusawithana, Maitri P and Cen, Cheng and Sleasman, Charles R and Woicik, Joseph C and Li, Yulan and Kourkoutis, Lena Fitting and Klug, Jeffrey A and Li, Hao and Ryan, Philip and Wang, Li-Peng and others},
  journal={Science},
  volume={324},
  number={5925},
  pages={367--370},
  year={2009},
  publisher={American Association for the Advancement of Science}
}

@article{li2006phase,
  title={Phase transitions and domain structures in strained pseudocubic (100) \rm {SrTiO$_3$} thin films},
  author={Li, YL and Choudhury, S and Haeni, JH and Biegalski, MD and Vasudevarao, A and Sharan, A and Ma, HZ and Levy, J and Gopalan, Venkatraman and Trolier-McKinstry, S and others},
  journal={Phys. Rev. B},
  volume={73},
  number={18},
  pages={184112},
  year={2006},
  publisher={APS}
}

@article{vasudevarao2006multiferroic,
  title={Multiferroic domain dynamics in strained strontium titanate},
  author={Vasudevarao, A and Kumar, A and Tian, L and Haeni, JH and Li, YL and Eklund, C-J and Jia, QX and Uecker, R and Reiche, P and Rabe, KM and others},
  journal={Phys. Rev. Lett.},
  volume={97},
  number={25},
  pages={257602},
  year={2006},
  publisher={APS}
}

@article{berger2011band,
  title={Band Gap and Edge Engineering via Ferroic Distortion and Anisotropic Strain: The Case of \rm {SrTiO$_3$}},
  author={Berger, Robert F and Fennie, Craig J and Neaton, Jeffrey B},
  journal={Phys. Rev. Lett.},
  volume={107},
  number={14},
  pages={146804},
  year={2011},
  publisher={APS}
}

@article{zhang2024intrinsic,
  title={Intrinsic-strain-induced ferroelectric order and ultrafine nanodomains in \rm {SrTiO$_3$}},
  author={Zhang, Peixi and Li, Qiang and Li, Zhiguo and Shi, Xiaoming and Wang, Haoyu and Huo, Chuanrui and Zhou, Lihui and Kuang, Xiaojun and Lin, Kun and Cao, Yili and others},
  journal={Proc. Natl. Acad. Sci. U.S. A.},
  volume={121},
  number={25},
  pages={e2400568121},
  year={2024},
  publisher={National Academy of Sciences}
}

@article{devonshire1954theory,
  title={Theory of ferroelectrics},
  author={Devonshire, AF0056},
  journal={Adv. Phys.},
  volume={3},
  number={10},
  pages={85--130},
  year={1954},
  publisher={Taylor \& Francis}
}

@article{salje2012ferroelastic,
  title={Ferroelastic materials},
  author={Salje, Ekhard KH},
  journal={Annu. Rev. Mater. Res.},
  volume={42},
  number={1},
  pages={265--283},
  year={2012},
  publisher={Annual Reviews}
}

@article{moya2014caloric,
  title={Caloric materials near ferroic phase transitions},
  author={Moya, X and Kar-Narayan, Sohini and Mathur, Neil David},
  journal={Nat. Mater.},
  volume={13},
  number={5},
  pages={439--450},
  year={2014},
  publisher={Nature Publishing Group UK London}
}

@article{yang2018flexo,
  title={Flexo-photovoltaic effect},
  author={Yang, Ming-Min and Kim, Dong Jik and Alexe, Marin},
  journal={Science},
  volume={360},
  number={6391},
  pages={904--907},
  year={2018},
  publisher={American Association for the Advancement of Science}
}

@article{shu2020photoflexoelectric,
  title={Photoflexoelectric effect in halide perovskites},
  author={Shu, Longlong and Ke, Shanming and Fei, Linfeng and Huang, Wenbin and Wang, Zhiguo and Gong, Jinhui and Jiang, Xiaoning and Wang, Li and Li, Fei and Lei, Shuijin and others},
  journal={Nat. Mater.},
  volume={19},
  number={6},
  pages={605--609},
  year={2020},
  publisher={Nature Publishing Group UK London}
}

@article{grosso2021prediction,
  title={Prediction of low-energy phases of \rm {BiFeO$_3$} with large unit cells and complex tilts beyond Glazer notation},
  author={Grosso, Bastien F and Spaldin, Nicola A},
  journal={Phys. Rev. Mater.},
  volume={5},
  number={5},
  pages={054403},
  year={2021},
  publisher={APS}
}

@article{biancoli2015breaking,
  title={Breaking of macroscopic centric symmetry in paraelectric phases of ferroelectric materials and implications for flexoelectricity},
  author={Biancoli, Alberto and Fancher, Chris M and Jones, Jacob L and Damjanovic, Dragan},
  journal={Nat. Mater.},
  volume={14},
  number={2},
  pages={224--229},
  year={2015},
  publisher={Nature Publishing Group UK London}
}

@article{dai2012magnetism,
  title={Magnetism and its microscopic origin in iron-based high-temperature superconductors},
  author={Dai, Pengcheng and Hu, Jiangping and Dagotto, Elbio},
  journal={Nat. Phys.},
  volume={8},
  number={10},
  pages={709--718},
  year={2012},
  publisher={Nature Publishing Group UK London}
}

@article{orenstein2025observation,
  title={Observation of polarization density waves in \rm {SrTiO$_3$}},
  author={Orenstein, Gal and Krapivin, Viktor and Huang, Yijing and Zhang, Zhuquan and de la Pe{\~n}a Mu{\~n}oz, Gilberto and Duncan, Ryan A and Nguyen, Quynh and Stanton, Jade and Teitelbaum, Samuel and Yavas, Hasan and others},
  journal={Nat. Phys.},
  pages={1--5},
  year={2025},
  publisher={Nature Publishing Group UK London}
}

@article{han2024freestanding,
  title={Freestanding perovskite oxide membranes: a new playground for novel ferroic properties and applications},
  author={Han, Lu and Dong, Guohua and Liu, Ming and Nie, Yuefeng},
  journal={Adv. Funct. Mater.},
  volume={34},
  number={4},
  pages={2309543},
  year={2024},
  publisher={Wiley Online Library}
}

@article{li2025classical,
  title={The classical-to-quantum crossover in the strain-induced ferroelectric transition in SrTiO3 membranes},
  author={Li, Jiarui and Lee, Yonghun and Choi, Yongseong and Kim, Jong-Woo and Thompson, Paul and Crust, Kevin J and Xu, Ruijuan and Hwang, Harold Y and Ryan, Philip J and Lee, Wei-Sheng},
  journal={Nat. Commun.},
  volume={16},
  number={1},
  pages={4445},
  year={2025},
  publisher={Nature Publishing Group UK London}
}

@article{chen2022resistive,
  title={Resistive Memory Based on the Spin-Density-Wave Transition of Antiferromagnetic Chromium},
  author={Chen, Hongyu and Feng, Zexin and Qin, Peixin and Zhou, Xiaorong and Yan, Han and Wang, Xiaoning and Meng, Ziang and Liu, Li and Liu, Zhiqi},
  journal={Phys. Rev. Appl.},
  volume={18},
  number={5},
  pages={054046},
  year={2022},
  publisher={APS}
}

@article{perdew1981self,
  title={Self-interaction correction to density-functional approximations for many-electron systems},
  author={Perdew, John P and Zunger, Alex},
  journal={Phys. Rev. B},
  volume={23},
  number={10},
  pages={5048},
  year={1981},
  publisher={APS}
}

@article{junquera2023topological,
  title={Topological phases in polar oxide nanostructures},
  author={Junquera, Javier and Nahas, Yousra and Prokhorenko, Sergei and Bellaiche, Laurent and {\'I}{\~n}iguez, Jorge and Schlom, Darrell G and Chen, Long-Qing and Salahuddin, Sayeef and Muller, David A and Martin, Lane W and others},
  journal={Rev. Mod. Phys.},
  volume={95},
  number={2},
  pages={025001},
  year={2023},
  publisher={APS}
}

@article{gruner1994dynamics,
  title={The dynamics of spin-density waves},
  author={Gr{\"u}ner, G},
  journal={Rev. Mod. Phys.},
  volume={66},
  number={1},
  pages={1},
  year={1994},
  publisher={APS}
}

@article{enderlein2020superconductivity,
  title={Superconductivity mediated by polar modes in ferroelectric metals},
  author={Enderlein, C and de Oliveira, J Ferreira and Tompsett, DA and Saitovitch, E Baggio and Saxena, SS and Lonzarich, GG and Rowley, SE},
  journal={Nat. Commun.},
  volume={11},
  number={1},
  pages={4852},
  year={2020},
  publisher={Nature Publishing Group UK London}
}

@article{li2017thickness,
  title={Thickness-dependent $a_1/a_2$ domain evolution in ferroelectric \rm {PbTiO$_3$} films},
  author={Li, S and Zhu, YL and Tang, Y Liu and Liu, Y and Zhang, SR and Wang, YJ and Ma, XL},
  journal={Acta Mater.},
  volume={131},
  pages={123--130},
  year={2017},
  publisher={Elsevier}
}

@MISC{Note1,note="see Supplemental Material, which includes Refs. [46-
57], computational details of DFT, Modulated ground state for SrMnO$_3$; Potential energy surface of individual and coupled modes; Polarization dependent magnetization deviation; Strain and polarization fields in lower-energy modulated BiFeO$_3$ and PbTiO$_3$."}

@article{stokes2006isodisplace,
  title={ISODISPLACE: a web-based tool for exploring structural distortions},
  author={Stokes, Harold T and Hatch, Dorian M and Campbell, Branton J and Tanner, David E},
  journal={ J. Appl. Crystallogr.},
  volume={39},
  number={4},
  pages={607--614},
  year={2006},
  publisher={Wiley Online Library}
}

@article{lee2010epitaxial,
  title={Epitaxial-strain-induced multiferroicity in \rm {SrMnO$_3$} from first principles},
  author={Lee, Jun Hee and Rabe, Karin M},
  journal={Phys. Rev. Lett.},
  volume={104},
  number={20},
  pages={207204},
  year={2010},
  publisher={APS}
}

@article{cazorla2023giant,
  title={Giant multiphononic effects in a perovskite oxide},
  author={Cazorla, Claudio and Stengel, Massimiliano and {\'I}{\~n}iguez, Jorge and Rurali, Riccardo},
  journal={Npj Comput. Mater.},
  volume={9},
  number={1},
  pages={97},
  year={2023},
  publisher={Nature Publishing Group UK London}
}

@article{guo2018strain,
  title={Strain-induced ferroelectricity and spin-lattice coupling in \rm {SrMnO$_3$} thin films},
  author={Guo, JW and Wang, PS and Yuan, Y and He, Q and Lu, JL and Chen, TZ and Yang, SZ and Wang, YJ and Erni, R and Rossell, MD and others},
  journal={Phys. Rev. B},
  volume={97},
  number={23},
  pages={235135},
  year={2018},
  publisher={APS}
}

@article{zubko2007strain,
  title={Strain-gradient-induced polarization in SrTiO 3 single crystals},
  author={Zubko, P and Catalan, G and Buckley, A and Welche, PRL and Scott, JF},
  journal={Phys. Rev. Lett.},
  volume={99},
  number={16},
  pages={167601},
  year={2007},
  publisher={APS}
}

@article{schiaffino2017macroscopic,
  title={Macroscopic polarization from antiferrodistortive cycloids in ferroelastic SrTiO 3},
  author={Schiaffino, Andrea and Stengel, Massimiliano},
  journal={Phys. Rev. Lett.},
  volume={119},
  number={13},
  pages={137601},
  year={2017},
  publisher={APS}
}

@article{liechtenstein1995density,
  title={Density-functional theory and strong interactions: Orbital ordering in Mott-Hubbard insulators},
  author={Liechtenstein, AI and Anisimov, Vladimir I and Zaanen, Jan},
  journal={Phys. Rev. B},
  volume={52},
  number={8},
  pages={R5467},
  year={1995},
  publisher={APS}
}

@article{kresse1993ab,
  title={Ab initio molecular dynamics for liquid metals},
  author={Kresse, Georg and Hafner, J{\"u}rgen},
  journal={Phys. Rev. B},
  volume={47},
  number={1},
  pages={558},
  year={1993},
  publisher={APS}
}

@article{blochl1994projector,
  title={Projector augmented-wave method},
  author={Bl{\"o}chl, Peter E},
  journal={Phys. Rev. B},
  volume={50},
  number={24},
  pages={17953},
  year={1994},
  publisher={APS}
}

@article{wang2020converse,
  title={Converse flexoelectricity around ferroelectric domain walls},
  author={Wang, YJ and Tang, YL and Zhu, YL and Feng, YP and Ma, XL},
  journal={Acta Mater.},
  volume={191},
  pages={158--165},
  year={2020},
  publisher={Elsevier}
}

@article{togo2015first,
  title={First principles phonon calculations in materials science},
  author={Togo, Atsushi and Tanaka, Isao},
  journal={Scr. Mater.},
  volume={108},
  pages={1--5},
  year={2015},
  publisher={Elsevier}
}

@article{makushko2022flexomagnetism,
  title={Flexomagnetism and vertically graded N{\'e}el temperature of antiferromagnetic \rm {$Cr_2O_3$} thin films},
  author={Makushko, Pavlo and Kosub, Tobias and Pylypovskyi, Oleksandr V and Hedrich, Natascha and Li, Jiang and Pashkin, Alexej and Avdoshenko, Stanislav and H{\"u}bner, Ren{\'e} and Ganss, Fabian and Wolf, Daniel and others},
  journal={Nat. Commun.},
  volume={13},
  number={1},
  pages={6745},
  year={2022},
  publisher={Nature Publishing Group UK London}
}

@article{yao2025fluctuating,
  title={Fluctuating local polarization: a generic fingerprint for enhanced piezoelectricity in Pb-based and Pb-free perovskite ferroelectrics},
  author={Yao, Yonghao and Liu, Hui and Hu, Yihao and Datta, Kaustuv and Wu, Jiagang and Zhang, Yuanpeng and Tucker, Matthew G and Liu, Shi and Neuefeind, Joerg C and Zhang, Shujun and others},
  journal={Nat. Commun.},
  volume={16},
  number={1},
  pages={7442},
  year={2025},
  publisher={Nature Publishing Group UK London}
}

@article{belyaev2020strain,
  title={Strain-Gradient-Induced Unidirectional Magnetic Anisotropy in Nanocrystalline Thin Permalloy Films},
  author={Belyaev, Boris A and Izotov, Andrey V and Solovev, Platon N and Boev, Nikita M},
  journal={Phys. Status Solidi. RRL},
  volume={14},
  number={1},
  pages={1900467},
  year={2020},
  publisher={Wiley Online Library}
}

@article{cowley1976acoustic,
  title={Acoustic phonon instabilities and structural phase transitions},
  author={Cowley, RA},
  journal={Phys. Rev. B},
  volume={13},
  number={11},
  pages={4877},
  year={1976},
  publisher={APS}
}

@article{schmid1994multi,
  title={Multi-ferroic magnetoelectrics},
  author={Schmid, Hans},
  journal={Ferroelectrics},
  volume={162},
  number={1},
  pages={317--338},
  year={1994},
  publisher={Taylor \& Francis}
}

@article{fiebig2016evolution,
  title={The evolution of multiferroics},
  author={Fiebig, Manfred and Lottermoser, Thomas and Meier, Dennis and Trassin, Morgan},
  journal={Nat. Rev. Mater.},
  volume={1},
  number={8},
  pages={1--14},
  year={2016},
  publisher={Nature Publishing Group}
}

@article{eerenstein2006multiferroic,
  title={Multiferroic and magnetoelectric materials},
  author={Eerenstein, Wilma and Mathur, ND and Scott, James F},
  journal={Nature},
  volume={442},
  number={7104},
  pages={759--765},
  year={2006},
  publisher={Nature Publishing Group UK London}
}

@article{martin2016thin,
  title={Thin-film ferroelectric materials and their applications},
  author={Martin, Lane W and Rappe, Andrew M},
  journal={Nat. Rev. Mater.},
  volume={2},
  number={2},
  pages={1--14},
  year={2016},
  publisher={Nature Publishing Group}
}

@article{muller1991indication,
  title={Indication for a novel phase in the quantum paraelectric regime of SrTiO3},
  author={M{\"u}ller, K Alex and Berlinger, W and Tosatti, E},
  journal={Z. Phys. B},
  volume={84},
  number={2},
  pages={277--283},
  year={1991},
  publisher={Springer}
}

@article{rowley2014ferroelectric,
  title={Ferroelectric quantum criticality},
  author={Rowley, SE and Spalek, LJ and Smith, RP and Dean, MPM and Itoh, M and Scott, JF and Lonzarich, GG and Saxena, SS},
  journal={Nat. Phys.},
  volume={10},
  number={5},
  pages={367--372},
  year={2014},
  publisher={Nature Publishing Group UK London}
}

@article{salje1990phase,
  title={Phase transitions in ferroelastic and co-elastic crystals},
  author={Salje, Ekhard},
  journal={Ferroelectrics},
  volume={104},
  number={1},
  pages={111--120},
  year={1990},
  publisher={Taylor \& Francis}
}
\end{document}